\documentclass[english,nofootinbib,amsmath,amssymb,superscriptaddress,preprint]{revtex4-2}
\usepackage[T1]{fontenc}
\usepackage[latin9]{inputenc}
\usepackage{color}
\usepackage{babel}
\usepackage{amsmath}
\usepackage{amssymb}
\usepackage{graphicx}
\usepackage[pdfusetitle,
 bookmarks=true,bookmarksnumbered=false,bookmarksopen=false,
 breaklinks=false,pdfborder={0 0 1},backref=false,colorlinks=true]
 {hyperref}
\hypersetup{
 linkcolor=blue,citecolor = blue }
\begin{document}
\global\long\def\tr{{\rm Tr}}%

\global\long\def\diag{{\rm diag}}%

\global\long\def\rad{{\rm rad}}%

\global\long\def\new{{\rm new}}%

\global\long\def\bp{{\rm \bar{\phi}}}%

\global\long\def\inta{{\rm int}}%

\global\long\def\ins{{\rm in}}%

\global\long\def\outs{{\rm out}}%

\global\long\def\re{{\rm Re}}%

\global\long\def\im{{\rm Im}}%

\title{Reconsidering the calculation of the false vacuum decay rate at zero temperature}
\author{Qiang Yin}
\affiliation{Graduate School of Science, Kyushu University, in.kyo@phys.kyushu-u.ac.jp}
\begin{abstract}
In the calculation of the decay rate at finite temperature using the saddle point approximation, we identified some inconsistencies in the calculation of the decay rate at zero temperature. These inconsistencies may impact the explanation provided by Callan and Coleman. To address these inconsistencies, we recalculated the decay rate using the shifted-bounce solution and the shot solution.
\end{abstract}
\maketitle
\tableofcontents{}

\section{Introduction}

In order to calculate the decay rate of the false vacuum state, Callan
and Coleman put forward their method in \citep{Coleman1977,Callan1977}.
They used the relation between the decay rate of a metastable state
and the corresponding imaginary part of its energy. They derived the
imaginary part of the energy with the help of path integral formalism
of the Euclidean transition amplitude. Then they obtained the final
result after applying the saddle point approximation.

However, there are some ambiguities in the choice of the classical
solutions and in the timing of taking the time interval to infinity.
These ambiguities caused some problems, especially in the application
of the collective coordinate method. Andreassen et al. tried to solve
some problems in their paper \citep{Andreassen2017} but there are
more. We would like to describe related problems first and try to
solve them using the shot solution presented in \citep{Andreassen2017}.

\section{Review of the decay rate at zero temperature\label{sec:review}}

The central idea for deriving the decay rate is the following relation
: 
\begin{equation}
\sum_{n}e^{-\frac{1}{\hbar}E_{n}{\cal T}}\langle x_{f}|n\rangle\langle n|x_{i}\rangle=\langle x_{f}|^{-\frac{1}{\hbar}\hat{H}{\cal T}}|x_{i}\rangle={\cal N}\int_{x(-\frac{{\cal T}}{2})=x_{i}}^{x(\frac{{\cal T}}{2})=x_{f}}{\cal D}xe^{-\frac{1}{\hbar}S_{E}[x]}.\label{eq:Trans-Path}
\end{equation}
By taking the Euclidean time ${\cal T}$ to infinity, the following
relation can be obtained : 
\begin{equation}
E_{0}=\lim_{{\cal T}\rightarrow\infty}-\frac{\hbar}{{\cal T}}\ln\langle x_{f}|^{-\frac{1}{\hbar}\hat{H}{\cal T}}|x_{i}\rangle,
\end{equation}
where $E_{0}$ is the energy of the state with the lowest energy.
Since the decay process of a metastable state is considered, $E_{0}$
should be the energy of the metastable state with the smallest real
part and the completeness relation used in Eq.(\ref{eq:Trans-Path})
should be the completeness relation of the metastable states. The
corresponding metastable state is called the false vacuum state. The
decay rate corresponding to the metastable state is correlated with
to the imaginary part of the complex energy as
\begin{equation}
\Gamma=-\frac{2}{\hbar}\im E_{0}.
\end{equation}

Since the Euclidean transition amplitude can be expressed as the path
integral formalism, the decay rate can be derived from the path integral
formalism directly. Moreover, the calculation of the path integral
can be performed by saddle point approximation.

\begin{figure}
\begin{centering}
\includegraphics[width=0.6\linewidth]{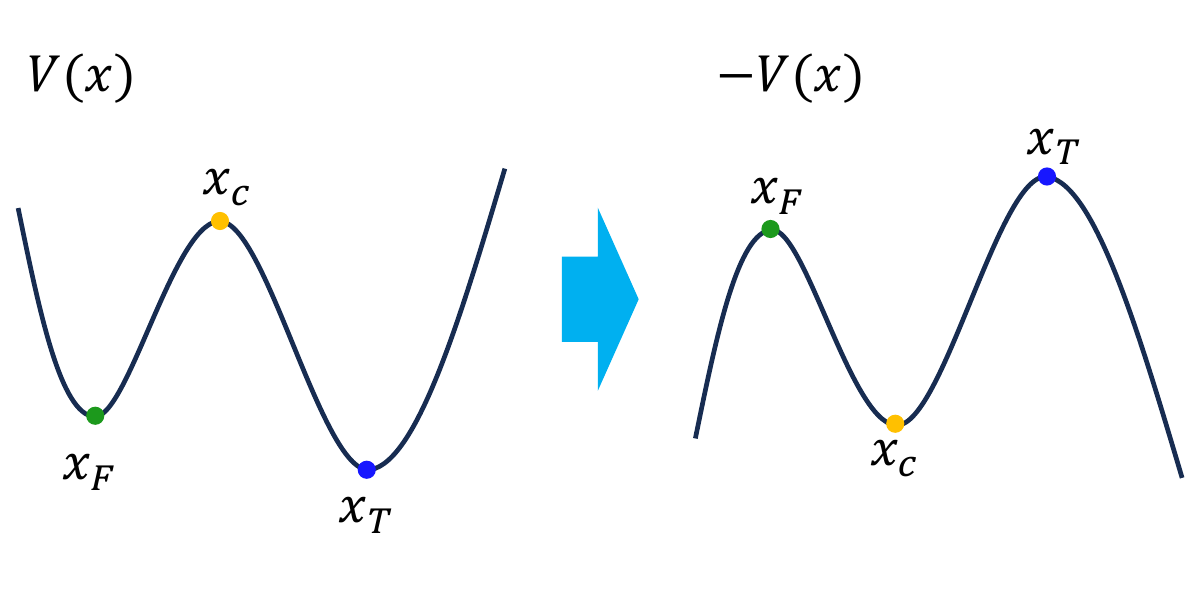}
\par\end{centering}
\caption{Potential and Euclidean Potential\label{fig:Potential-of-the-system}}
\end{figure}

Consider a system with the potential shown in Figure.\ref{fig:Potential-of-the-system}.
The decay process of the false vacuum state could occur in such a
system. As the Euclidean transition amplitude is adopted, the potential
in $S_{E}[x]$ is reversed. Although the selection of the endpoints
$x_{i}$ and $x_{f}$ is arbitrary, it is normally to choose them
to be $x_{i}=x_{f}=x_{F}$ in most of the research. Therefore, a classical
solution $\bar{x}(\tau)$ is required to satisfy 
\begin{equation}
\bar{x}(+\frac{{\cal T}}{2})=\bar{x}(-\frac{{\cal T}}{2})=x_{F}.
\end{equation}
When ${\cal T}$ is infinite, there are two classical solutions according
to Callan and Coleman. One is called the false vacuum solution $x_{F}(\tau)=x_{F}$.
The other is called the bounce $x_{B}(\tau)$.
\begin{figure}
\begin{centering}
\includegraphics[width=0.4\linewidth]{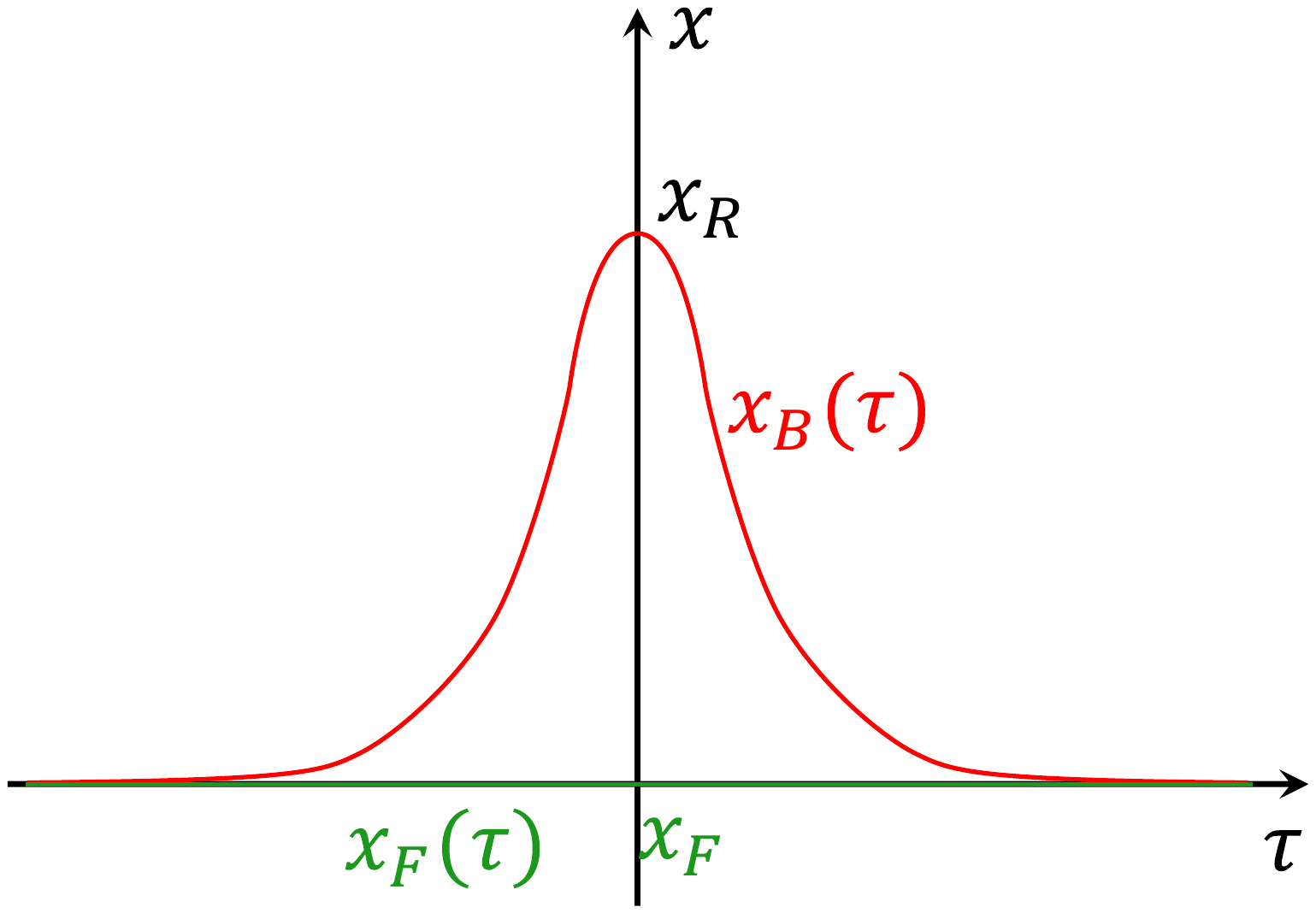}
\par\end{centering}
\caption{$x_{F}(\tau)$ and $x_{B}(\tau)$\label{fig:Potential-of-the-system-1}}
\end{figure}

After finding a classical solution $\bar{x}(\tau)$, the path $x(\tau)$
in the path integral can be expanded around $\bar{x}(\tau)$ as
\begin{equation}
x(\tau)=\bar{x}(\tau)+\Delta x(\tau).
\end{equation}
Then the exponent $S_{E}[x]$ becomes
\begin{equation}
S_{E}[x]=S_{E}[\bar{x}]+\frac{m}{2}\int d\tau\Delta x(\tau)\left(-\partial_{\tau}^{2}+\frac{1}{m}V^{\prime\prime}(\bar{x}(\tau))\right)\Delta x(\tau)
\end{equation}
at ${\cal O}(\hbar^{2})$. The fluctuation $\Delta x(\tau)$ can be
expanded by the eigenfunctions $x_{n}(\tau)$ of the fluctuation operator
$\left[-\partial^{2}+\frac{1}{m}V^{\prime\prime}(\bar{x}(\tau))\right]$
corresponding to the saddle point $\bar{x}(\tau)$. $x_{n}(\tau)$
satisfies the eigenequation
\begin{equation}
\left[-\partial^{2}+\frac{1}{m}V^{\prime\prime}(\bar{x}(\tau))\right]x_{n}(\tau)=\lambda_{n}x_{n}(\tau).
\end{equation}
Furthermore, since $\bar{x}(\tau)$ is a classical solution, it should
satisfy the same boundary condition as $x(\tau)$. As a result, $\Delta x$
and all $x_{n}(\tau)$s should satisfy the Dirichlet boundary conditions
:
\begin{equation}
\Delta x(+\frac{{\cal T}}{2})=\Delta x(-\frac{{\cal T}}{2})=0
\end{equation}
\begin{equation}
x_{n}(+\frac{{\cal T}}{2})=x_{n}(-\frac{{\cal T}}{2})=0
\end{equation}
in order to be in consistency. Then the expansion around $x(\tau)$
can be expressed as
\begin{equation}
x(\tau)=\bar{x}(\tau)+\Delta x=\bar{x}(\tau)+\sum_{n}\frac{c_{n}}{\sqrt{m}}x_{n}(\tau).
\end{equation}
And the contribution of $\bar{x}$ to the original path integral can
be expressed as
\begin{align}
{\cal N}\int{\cal D}xe^{-\frac{1}{\hbar}S_{E}[x]} & \approx{\cal N}\int{\cal D}\Delta x(\tau)e^{-\frac{1}{\hbar}\left(S_{E}[\bar{x}]+\frac{m}{2}\int d\tau\Delta x(\tau)\left(-\partial_{\tau}^{2}+\frac{1}{m}V^{\prime\prime}(\bar{x}(\tau))\right)\Delta x(\tau)\right)}\\
 & =e^{-\frac{1}{\hbar}S_{E}[\bar{x}]}{\cal N}\prod_{n}\int\frac{1}{\sqrt{2\pi\hbar}}e^{-\frac{1}{\hbar}\frac{1}{2}\lambda_{n}c_{n}^{2}}\\
 & =e^{-\frac{1}{\hbar}S_{E}[\bar{x}]}{\cal N}\prod_{n}\frac{1}{\sqrt{\lambda_{n}}}\\
 & ={\cal N}\frac{e^{-\frac{1}{\hbar}S_{E}[\bar{x}]}}{\sqrt{\det\left[-\partial^{2}+\frac{1}{m}V^{\prime\prime}(\bar{x}(\tau))\right]}}.
\end{align}

For finite ${\cal T}$, the contribution from $x_{F}(\tau)$ can be
expressed as
\begin{equation}
Z_{F}={\cal N}\int{\cal D}\Delta xe^{-\frac{1}{\hbar}\frac{m}{2}\int d\tau\Delta x(\tau)\left(-\partial_{\tau}^{2}+\omega_{F}^{2}\right)\Delta x(\tau)}={\cal N}\frac{1}{\sqrt{\det\left[-\partial^{2}+\omega_{F}^{2}\right]}}=\sqrt{\frac{\omega_{F}{\cal T}}{\sinh\left(\omega_{F}{\cal T}\right)}},
\end{equation}
where $\omega_{F}\equiv\sqrt{\frac{V^{\prime\prime}(x_{F})}{m}}$
and $V(x_{F})$ is set to zero.

As for the bounce $x_{B}(\tau)$ for infinite ${\cal T}$, its time
derivative satisfies 
\begin{equation}
\left[-\partial^{2}+\frac{1}{m}V^{\prime\prime}(x_{B}(\tau))\right]\dot{x}_{B}(\tau)=0,
\end{equation}
and $\dot{x}_{B}(+\frac{{\cal T}}{2})=\dot{x}_{B}(-\frac{{\cal T}}{2})=0$.
Consequently, it corresponds to an eigenfunction with a zero eigenvalue.
Moreover, $\dot{x}_{B}(\tau)$ has a node. Therefore, a negative
eigenvalue also exists. As a result, the integral over $c_{0}$(corresponding
to the negative mode $\lambda_{0}<0$) as well as $c_{1}$(corresponding
to the zero mode $\lambda_{1}=0$) becomes infinite.

The infinity caused by $\int\frac{1}{\sqrt{2\pi\hbar}}e^{-\frac{1}{\hbar}\frac{1}{2}\lambda_{0}c_{0}^{2}}$
was treated with analytic continuation. The negative eigenvalue indicates
that the bounce solution is not a minimum of $S_{E}[x]$ but a maximum
along this special direction, which corresponds to $x_{0}(\tau)$,
in the function space. Moreover, this direction was assumed to be
attached to one steepest ascent direction from the false vacuum solution.
By integrating over a series of paths $\widetilde{x}(\tau;z)$ parameterized
by a real variable $z$ along the direction, the path integral can
be written as 
\begin{equation}
Z_{\{z\}}=\int\frac{dz}{\sqrt{2\pi\hbar}}e^{-\frac{1}{\hbar}S_{E}[z]}.\label{eq:Z=00007Bz=00007D}
\end{equation}
$\widetilde{x}(\tau;z)$ should satisfy $\frac{d\widetilde{x}(\tau;z)}{dz}\bigg|_{z=1}=x_{0}(\tau)$
and can be set to equal to $x_{F}(\tau)$ at $z=0$ and $x_{B}(\tau)$
at $z=1$ separately. 
\begin{figure}
\begin{centering}
\includegraphics[width=0.4\linewidth]{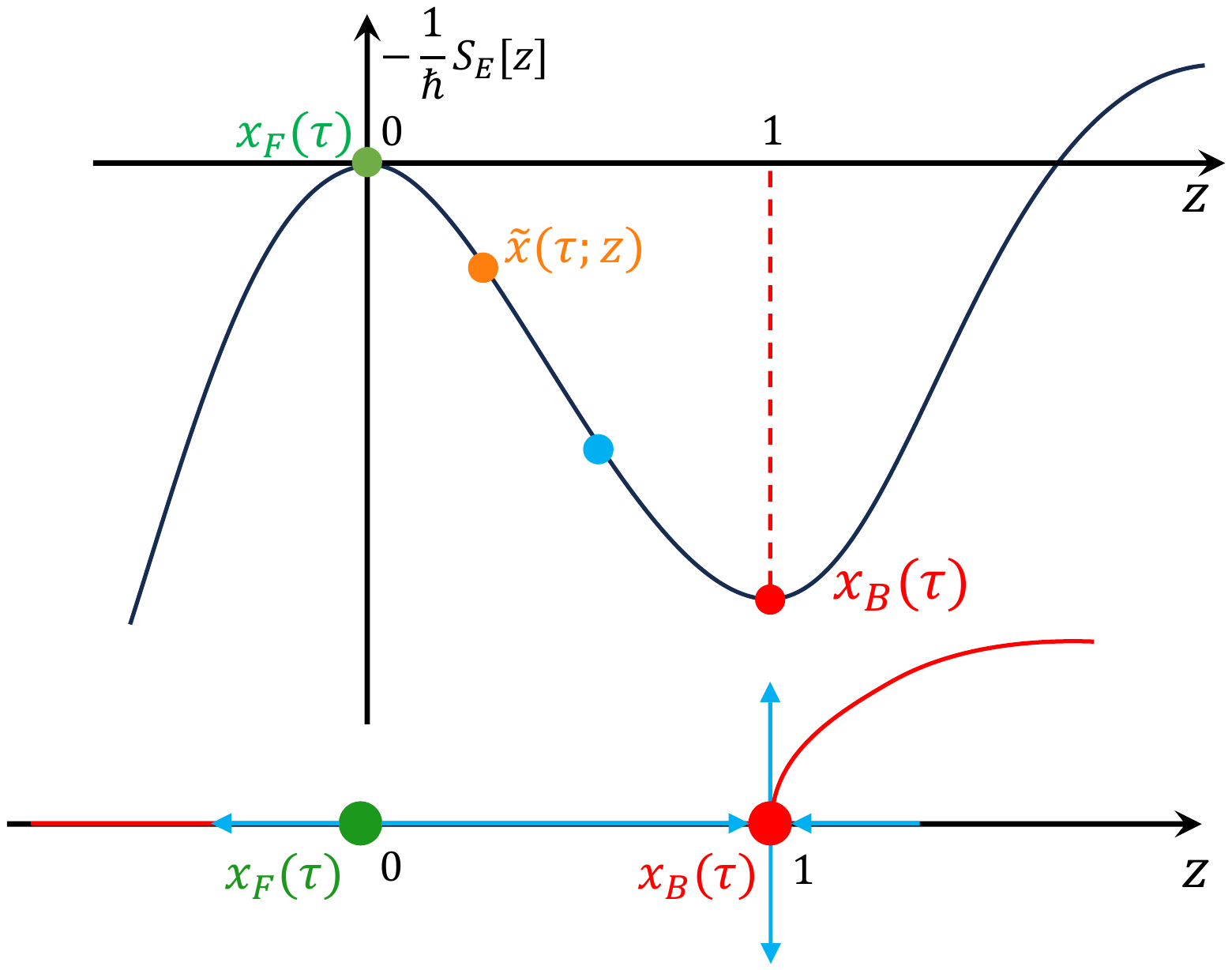}
\par\end{centering}
\caption{$-\frac{1}{\hbar}S_{E}[z]$}
\end{figure}
$S_{E}[z]$ are assumed to reach minus infinity when $z$ becomes
large enough. The divergence of $-\frac{1}{\hbar}S_{E}[z]$ will lead
to the divergence of $Z_{\{z\}}$, which corresponds to the divergence
of the integral over $c_{0}$. Hence, in order to obtain a reasonable
result, analytic continuation is required. Actually, the analytic
continuation can be conducted by changing the integral contour from
the real $z$-axis to a new integral contour, which was distorted
along the imaginary axis at $z=1$. The new contour is the steepest
descent contour passing through $x_{F}(\tau)$ following the previous
assumption. Along the new contour, the stationary point approximation
becomes
\begin{equation}
Z_{\{z\}}\approx Z_{z=0}+\frac{1}{2}Z_{z=1}\approx\frac{e^{-\frac{1}{\hbar}S_{E}[0]}}{\sqrt{S_{E}^{\prime\prime}[0]}}+\frac{1}{2}i\frac{e^{-\frac{1}{\hbar}S_{E}[1]}}{\sqrt{\left|S_{E}^{\prime\prime}[1]\right|}}.
\end{equation}

The infinity caused by $\int\frac{1}{\sqrt{2\pi\hbar}}e^{-\frac{1}{\hbar}\frac{1}{2}\lambda_{1}c_{1}^{2}}$
was treated with the so-called collective coordinate method. According
to \citep{Callan1977}, the relation between the expansion coefficient
$c_{1}$ of the zero eigenfunction and the time translations of the
bounce solution $\tau_{c}$ is given by
\begin{equation}
\frac{dc_{1}}{\sqrt{m}}\frac{\dot{x}_{B}(\tau)}{\left\Vert \dot{x}_{B}(\tau)\right\Vert }=d\tau_{c}
\end{equation}
since the small variation over $c_{1}$ can be expressed as
\begin{align}
dx(\tau) & =\frac{dc_{1}}{\sqrt{m}}x_{1}(\tau)=\frac{dc_{1}}{\sqrt{m}}\frac{\dot{x}_{B}(\tau)}{\left\Vert \dot{x}_{B}(\tau)\right\Vert }=x_{B}(\tau+\frac{dc_{1}}{\sqrt{m}\left\Vert \dot{x}_{B}(\tau)\right\Vert })-x_{B}(\tau).
\end{align}
The relation indicates that the small variation over $c_{1}$ can
be transformed into the small variation over $\tau_{c}$. By use of
$x_{1}(\tau)\frac{dc_{1}}{\sqrt{m}}=\dot{x}_{B}(\tau)d\tau_{c}$ and
$\frac{1}{2}\dot{x}_{B}^{2}(\tau)-V(x_{B}(\tau))=-V(x_{F})=0$, the
following relation can be obtained
\begin{equation}
\frac{1}{\sqrt{2\pi\hbar}}dc_{1}=\sqrt{\frac{S_{B}}{2\pi\hbar}}d\tau_{c}.\label{eq:dc1=00003Ddtauc}
\end{equation}
The integration over $c_{1}$ is thought to be equivalent to the integrals
over $\tau_{c}$. As a result, it is concluded that there is no need
to include the zero eigenvalue but to include a factor of $\sqrt{\frac{S_{B}}{2\pi\hbar}}d\tau_{c}$
since the integral over the center of the bounce has already been
performed.

Combining the result above, the saddle point approximation can be
expressed
\begin{align}
{\cal N}\int{\cal D}xe^{-\frac{1}{\hbar}S_{E}[x]} & \sim Z_{F}+\frac{1}{2}Z_{B}\\
 & ={\cal N}\frac{1}{\sqrt{\det\left[-\partial^{2}+\frac{1}{m}V^{\prime\prime}(x_{F})\right]}}+i\frac{1}{2}{\cal N}\frac{\sqrt{\frac{S_{B}}{2\pi\hbar}}{\cal T}e^{-\frac{1}{\hbar}S_{B}}}{\sqrt{\left|\det^{\prime}\left[-\partial^{2}+\frac{1}{m}V^{\prime\prime}(x_{B})\right]\right|}}.
\end{align}
And the imaginary part of $E_{0}$ becomes
\begin{align}
\lim_{{\cal T}\rightarrow\infty}\im E_{0} & =-\frac{\hbar}{{\cal T}}\im\ln\left(Z_{F}+\frac{1}{2}Z_{B}\right)\approx-\frac{\hbar}{{\cal T}}\im\frac{1}{2}\frac{Z_{B}}{Z_{F}}\\
 & \approx-\frac{\hbar}{2}\sqrt{\frac{S_{B}}{2\pi\hbar}}\sqrt{\frac{\det\left[-\partial^{2}+\frac{1}{m}V^{\prime\prime}(x_{F})\right]}{\left|\det^{\prime}\left[-\partial^{2}+\frac{1}{m}V^{\prime\prime}(x_{B})\right]\right|}}e^{-\frac{1}{\hbar}S_{B}},
\end{align}
where $\det^{\prime}$ means excluding the zero eigenvalue. The decay
rate at zero temperature then becomes
\begin{equation}
\Gamma=\sqrt{\frac{S_{B}}{2\pi\hbar}}\sqrt{\frac{\det\left[-\partial^{2}+\frac{1}{m}V^{\prime\prime}(x_{F})\right]}{\left|\det^{\prime}\left[-\partial^{2}+\frac{1}{m}V^{\prime\prime}(x_{B})\right]\right|}}e^{-\frac{1}{\hbar}S_{B}}.
\end{equation}

The ratio is defined at zero temperature and can be calculated following
\citep{Coleman1985} as 
\begin{equation}
\Gamma_{\textrm{C\&C}}=\sqrt{\frac{S_{B}}{2\pi\hbar}}\sqrt{2m\omega_{F}}A(\infty)e^{-\frac{1}{\hbar}S_{B}},
\end{equation}
where
\begin{equation}
A(\tau)\equiv\frac{\omega_{F}}{\sqrt{S_{B}}}\left(x_{R}-x_{F}\right)e^{\int_{x_{B}(\tau)}^{x_{R}}dx\left(\frac{\omega_{F}}{\sqrt{\frac{2}{m}\left(V(x)-V(x_{F})\right)}}-\frac{1}{|x-x_{F}|}\right)}.
\end{equation}

We should emphasize that the calculation of the ratio is carried out
under the condition that the time interval ${\cal T}$ is infinite.
Parts of the classical solutions in $[-\frac{{\cal P}}{2},\frac{{\cal P}}{2}]$
were extracted in order to perform a possible calculation. The corresponding
ratio related to these two partial classical solutions was calculated
first and then ${\cal P}$ was taken to infinity to obtain the final
result. This method is different from the method starting from a finite
time interval. And the ${\cal P}$ here is chosen artificially and
has nothing to do with the time interval ${\cal T}$.

\section{Reconsideration\label{sec:Reconsideration}}

There are at least two issues requiring further explanations in the
previous derivation. The first concerns the appearance of an imaginary
part in a Euclidean path integral. The second involves the timing
of taking ${\cal T}$ to infinity.

The first problem is somehow sophisticated. Part of the problem was
already solved in \citep{Andreassen2017}. Mathematically, the appearance
of the imaginary part originates from the particular selection of
the integral contour, or in other words, the selection of the saddle
points. Actually, another classical solution called the shot solution
$x_{S}(\tau)$ exists according to \citep{Andreassen2017}. The direction
related to the negative mode of the bounce is assumed to be attached
to the shot solution as well. Consequently, the shot solution will
be another minimum and $-\frac{1}{\hbar}S_{E}[z]$ won't reach minus
infinity as $z$ becomes large. Therefore, the one dimensional integral
Eq.(\ref{eq:Z=00007Bz=00007D}) won't diverge. In order to obtain
the imaginary part, the integral contour should be chosen as the steepest
descent contour passing through the false vacuum solution rather than
the original real axis, which excludes the contribution from the shot
solution and leads to the imaginary part. In other words, the imaginary
part appears because of the selection of a special integral contour
rather than a proper analytic continuation. 
\begin{figure}
\begin{centering}
\includegraphics[width=0.4\linewidth]{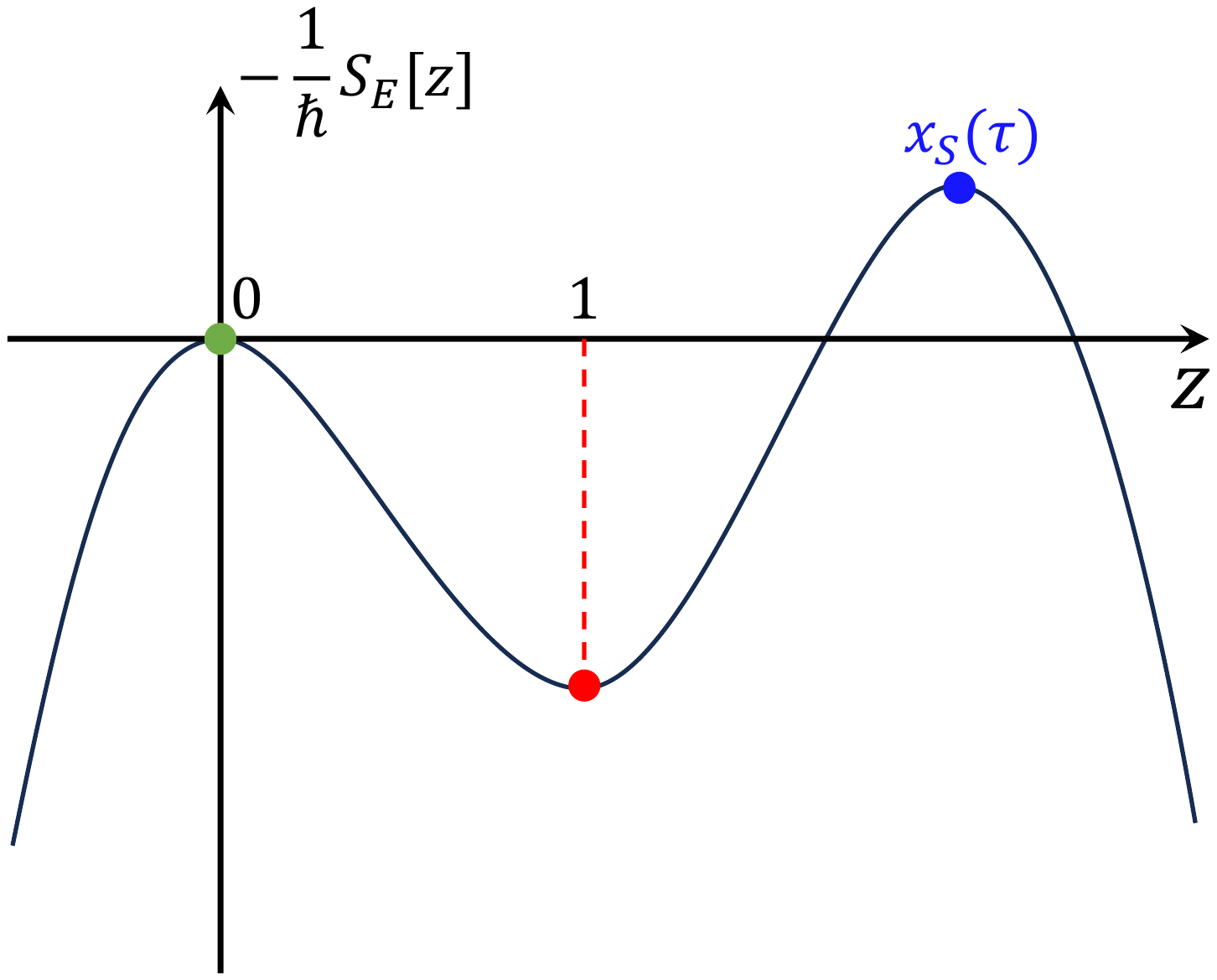}
\par\end{centering}
\caption{$-\frac{1}{\hbar}S_{E}[z]$}
\end{figure}

However, the reason for such a choice of contour was not fully explained
in \citep{Andreassen2017}, especially for the Euclidean time formalism.
We believe that the reason is related to the properties of the metastable
states since such states with complex energies should be defined by
particular boundary conditions which are different from the boundary
conditions of the bound states. Such boundary conditions should show
their influences apparently in the path integral, such as the choice
of the integral contour. To solve the problem completely, we believe
that some basic concepts such as the rigged Hilbert space (see \citep{Boehm1993})
\[
\Phi\subset{\cal H}\subset\Phi^{\times},
\]
which includes states with different boundary conditions such as Gamow
vectors, should be considered since both states with complex energy
and $|x\rangle$s used to generate the path integral formalism are
vectors in the rigged Hilbert space. However, we won't discuss the
problem in this paper.

The second problem was confused\textbf{ }at the beginning of the derivation
of Callan and Coleman. Although they started their discussion from
a finite time interval ${\cal T}$, they included the bounce solution
$x_{B}(\tau)$ with the endpoints at $x_{F}$, which can only exist
at an infinite time interval. If the contribution from the bounce
solution is included while calculating the path integral by saddle
point approximation, then all ${\cal T}$ appearing in the intermediate
process are merely symbols representing infinity rather than physical
quantities with physical meanings. This problem is closely related
to the validity of using the collective coordinate method.

Callan and Coleman used the collective coordinate method to deal with
the zero mode existing in the fluctuation determinant of the bounce
solution. The result of the zero mode should be $\int\frac{dc_{1}}{\sqrt{2\pi\hbar}}$,
which is definitely infinite. It can't be equal to $\int\sqrt{\frac{S_{B}}{2\pi\hbar}}d\tau_{c}$
if the integral range is finite. Therefore, it is impossible to use
a finite physical integral 
\begin{equation}
\int\sqrt{\frac{S_{B}}{2\pi\hbar}}d\tau_{c}=\sqrt{\frac{S_{B}}{2\pi\hbar}}{\cal T}
\end{equation}
to represent an infinite integral $\int\frac{dc_{1}}{\sqrt{2\pi\hbar}}$
unless the ${\cal T}$ here is just a symbol representing infinity.
In fact, as we discussed in our paper \citep{Harada2025}, the bounce
solution with different endpoints also exists when the time interval
is finite. Some may think that there is no zero mode of the fluctuation
determinant of such bounce solution and only a quasi-zero mode at
large time interval. Unfortunately, the fluctuation determinants of
those bounce solutions also have an exact zero eigenvalue no matter
what the time interval is since the initial velocities of these bounce
solutions are always zero. As a result, their time derivatives serve
as the eigenfunctions with zero eigenvalues, leading to divergence.
Therefore, it can't be expected that a smooth transition exists while
the time interval changes from finite to infinite since the zero mode
always exists.

According to Langer's explanation in \citep{Langer1967}, shifted-bounces
$x_{B}(\tau-\tau_{c})$ with different central values trace out a
line in the function space. All points on the line have completely
the same contributions. Consequently, the length of the line which
is equal to $\int\sqrt{\frac{S_{B}}{2\pi\hbar}}d\tau_{c}$ was calculated
instead of the zero mode. This explanation is similar to the explanation
given in \citep{Andreassen2017}. They tried to explain the inconsistencies
appeared in using collective coordinate method from the point of view
of the coordinate transformation. They changed the original integral
over 
\begin{equation}
\{c_{0},c_{1},\cdots\}:x(c_{0},c_{1},\cdots)=x_{B}(\tau)+\sum_{n}c_{n}x_{n}(\tau)
\end{equation}
to 
\begin{equation}
\{\tau_{c},\zeta_{0},\zeta_{2},\cdots\}:x(\tau_{c},\zeta_{0},\zeta_{2},\cdots)=x_{B}(\tau-\tau_{c})+\sum_{n\neq1}\zeta_{n}x_{n}(\tau-\tau_{c}).
\end{equation}

However, both of their explanations are also insufficient. First,
both explanations require that the original paths in the path integral
should satisfy the periodic boundary conditions instead of the Dirichlet
boundary condition. Otherwise, the boundary conditions of the original
path can't be satisfied for nonzero or large $\tau_{c}$.

Furthermore, the equivalence between the two coordinate systems is
suspicious. $\{c_{n}\}$ is a local coordinate system created around
a saddle point while $\tau_{c}$ is a global coordinate axis in function
space. They are definitely not equivalent for finite ${\cal T}$.
Whether they become the same or not at infinite ${\cal T}$ requires
further proof. For simplicity, whether the relation Eq.(\ref{eq:dc1=00003Ddtauc})
between $c_{1}$ and $\tau_{c}$ holds for all $c_{1}$ and $\tau_{c}$
requires further explanation.

\section{Starting from finite time interval}

In fact, if we admit the explanation for the first problem, we can
avoid the second problem in calculating the transition amplitude.
Although there is always a zero mode related to a bounce solution,
it is actually not necessary to include a bounce solution. The shifted-bounce
solution and the shot-F solution were used to calculate the free energy
at finite temperature in our last paper \citep{Harada2025}. They
also exist when the time interval is finite and can be used to calculate
the transition amplitude as well. Fortunately, the fluctuation operator
of a shifted-bounce solution $x_{sB}(\tau;\tau_{c})$ with nonzero
initial velocity has no zero eigenvalue. 
\begin{figure}[h]
\begin{centering}
\includegraphics[width=0.4\linewidth]{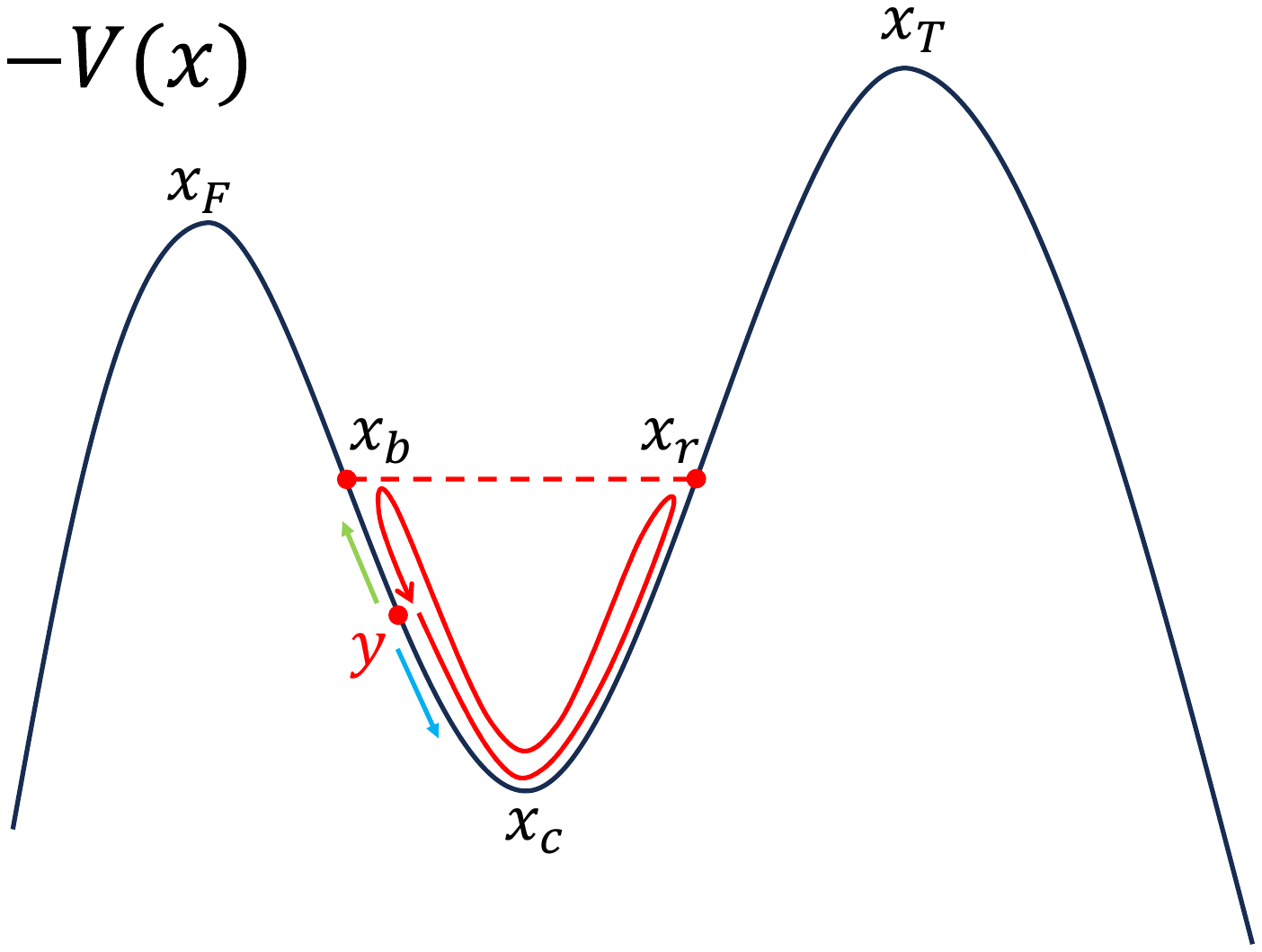}
\par\end{centering}
\caption{Shifted-bounce solution\label{fig:Shifted-bounce-solution}}
\end{figure}

Consider the finite time interval ${\cal T}$, the endpoints of a
bounce shift from $x_{F}$ to some other point $x_{b}$. Since $x_{i}$
and $x_{f}$ can be chosen arbitrarily, they can be chosen to be $y$,
which is different from $x_{b}$ as shown in Figure.\ref{fig:Shifted-bounce-solution}.
\begin{equation}
E_{0}=\lim_{{\cal T}\rightarrow\infty}-\frac{\hbar}{{\cal T}}\ln\langle x_{f}|^{-\frac{1}{\hbar}\hat{H}{\cal T}}|x_{i}\rangle=\lim_{{\cal T}\rightarrow\infty}-\frac{\hbar}{{\cal T}}\ln{\cal N}\int_{x(-\frac{{\cal T}}{2})=y}^{x(\frac{{\cal T}}{2})=y}{\cal D}xe^{-\frac{1}{\hbar}S_{E}[x]}.
\end{equation}
Consequently, the related classical solutions are a shifted-bounce
solution $x_{sB}^{{\cal T}}(\tau;y)$ starting from $y$ and the corresponding
shot-F solution $x_{sF}^{{\cal T}}(\tau)$. As long as the initial
velocity of the shifted-bounce solution is not zero, no zero mode
appears. We can calculate the transition amplitude from a ``real''
finite time interval ${\cal T}$ and then take ${\cal T}$ to infinity.
All ${\cal T}$ inside the limitation are finite. 
\begin{figure}[h]
\begin{centering}
\includegraphics[width=0.4\linewidth]{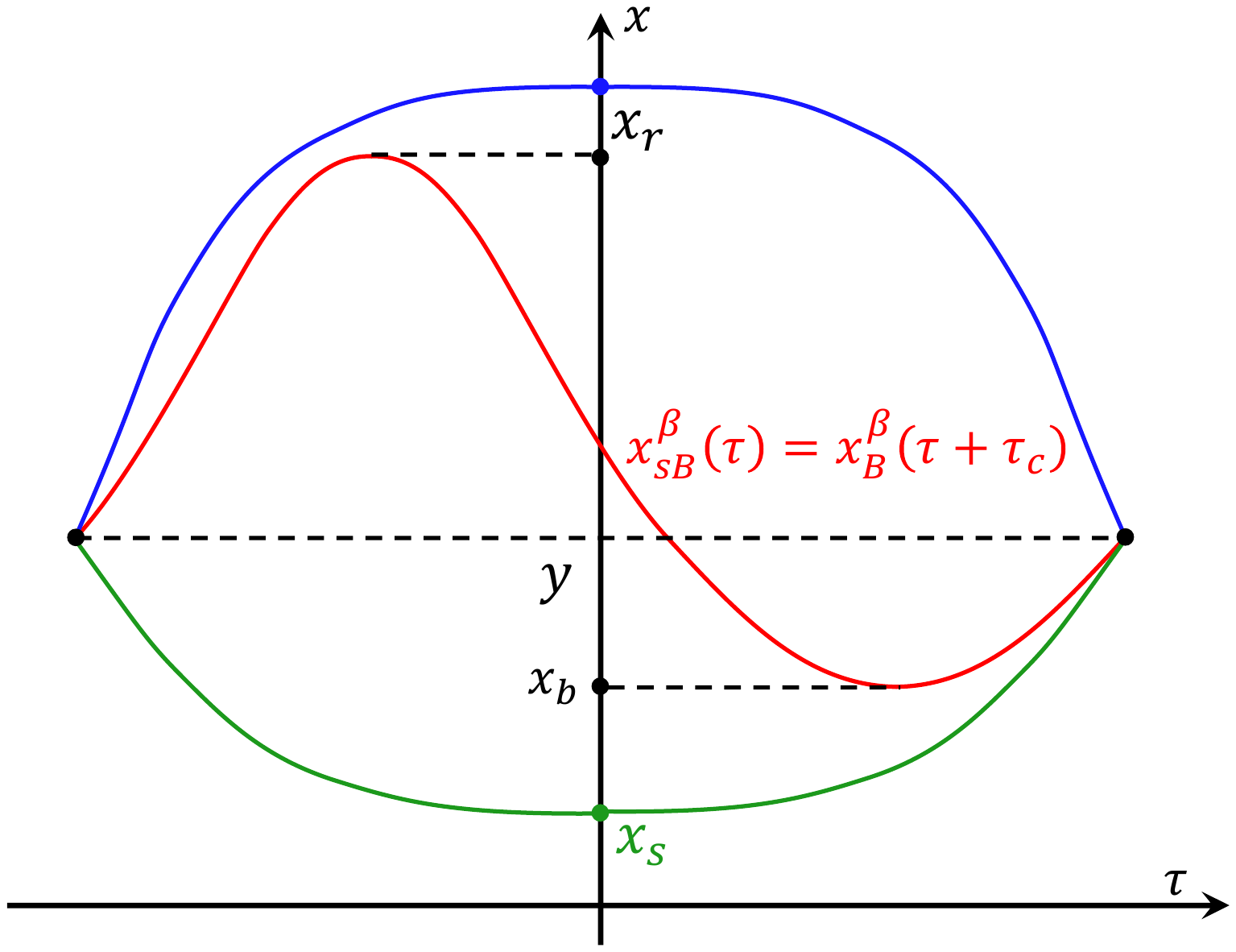}
\par\end{centering}
\caption{The bounce solution and corresponding shot solutions\label{fig:Classical solutions at finite temperature}}
\end{figure}

Although there could be several classical solutions, since we take
the integral contour to be the steepest descent contour passing through
$x_{sF}(\tau)$, only the contributions from the shifted-bounce solution
$x_{sB}(\tau)$ and the shot-F solution $x_{sF}(\tau)$ are contained.
As a result, the path integral can be expressed as
\begin{equation}
{\cal N}\int_{x(-\frac{{\cal T}}{2})=y}^{x(\frac{{\cal T}}{2})=y}{\cal D}xe^{-\frac{1}{\hbar}S_{E}[x]}=Z_{sF}^{{\cal T}}+\frac{1}{2}Z_{sB}^{{\cal T}},
\end{equation}
where the analytic continuation was performed and the contributions
from other classical solutions were excluded in order to obtain an
imaginary part\footnote{Actually, two shifted-bounce solutions with opposite initial velocities
exist. Their contributions are the same. However, our main purpose
focuses on the second issue mentioned in the previous section rather
than providing a complete calculation. Therefore, we won't discuss
the choice for the classical solutions here. }. 

Then the imaginary part of the energy can be expressed as
\begin{align}
\im E_{0} & =\lim_{{\cal T}\rightarrow\infty}-\frac{\hbar}{{\cal T}}\im\ln{\cal N}\int_{x(-\frac{{\cal T}}{2})=y}^{x(\frac{{\cal T}}{2})=y}{\cal D}xe^{-\frac{1}{\hbar}S_{E}[x]}\approx\lim_{{\cal T}\rightarrow\infty}-\frac{\hbar}{{\cal T}}\im\frac{\frac{1}{2}Z_{sB}^{{\cal T}}}{Z_{sF}^{{\cal T}}}\\
 & =\lim_{{\cal T}\rightarrow\infty}-\frac{\hbar}{2}\frac{1}{{\cal T}}\sqrt{\frac{\det\left[-\partial_{\tau}^{2}+\frac{1}{m}V^{\prime\prime}(x_{sF}^{{\cal T}}(\tau))\right]}{\left|\det\left[-\partial_{\tau}^{2}+\frac{1}{m}V^{\prime\prime}(x_{sB}^{{\cal T}}(\tau;y))\right]\right|}}e^{-\frac{1}{\hbar}\left[S_{sB}^{{\cal T}}-S_{sF}^{{\cal T}}\right]}.
\end{align}

The ratio can be calculated as
\begin{equation}
\frac{\det\left[-\partial_{\tau}^{2}+\frac{1}{m}V^{\prime\prime}(x_{sF}^{{\cal T}}(\tau))\right]}{\det\left[-\partial_{\tau}^{2}+\frac{1}{m}V^{\prime\prime}(x_{sB}^{{\cal T}}(\tau;y))\right]}=-\frac{-\frac{\partial S_{sF}}{\partial E_{sF}}}{m\left[\dot{x}_{sB}^{{\cal T}}(\frac{{\cal T}}{2};y)\right]^{2}\frac{d{\cal T}}{dE_{B}}},
\end{equation}
where $S_{sF}$ is the action of $x_{sF}^{{\cal T}}(\tau)$ and $E_{sF}$
is its energy. Denote the turning point of the shot-F solution as
$x_{s}$, then the ration can be expressed as
\begin{align}
\frac{\det\left[-\partial_{\tau}^{2}+\frac{1}{m}V^{\prime\prime}(x_{sF}^{{\cal T}}(\tau))\right]}{\det\left[-\partial_{\tau}^{2}+\frac{1}{m}V^{\prime\prime}(x_{sB}^{{\cal T}}(\tau;y))\right]} & =\frac{2\sqrt{2m\left(V(y)-V(x_{s})\right)}\frac{1}{V^{\prime}(x_{s})}\frac{dy}{dx_{s}}}{2\left(V(y)-V(x_{b})\right)\frac{1}{V^{\prime}(x_{b})}\frac{d{\cal T}}{dx_{b}}}\\
 & =\frac{2}{\sqrt{\frac{2}{m}\left(V(y)-V(x_{b})\right)}}\sqrt{\frac{V(y)-V(x_{s})}{V(y)-V(x_{b})}}\left[\frac{V^{\prime}(x_{b})}{V^{\prime}(x_{s})}\right]^{2}\frac{V^{\prime}(x_{s})\frac{dy}{dx_{s}}}{V^{\prime}(x_{b})\frac{d{\cal T}}{dx_{b}}}
\end{align}
When ${\cal T}\rightarrow\infty$, $x_{s}\leq x_{b}\rightarrow x_{F}$.
In order to obtain the final result, we need to find the limit relation
between ${\cal T}$, $x_{b}$ and $x_{s}$ following the method used
in \citep{Marino2015}.

We start from the derivation of the limit of $\frac{V^{\prime}(x_{b})}{V^{\prime}(x_{s})}$.
Here, since ${\cal T}$ is the time interval, it can be expressed
as the integral :
\begin{equation}
{\cal T}=2\int_{x_{b}}^{x_{r}}\frac{1}{\sqrt{\frac{2}{m}\left(V(x)-V(x_{b})\right)}}dx=2\int_{x_{s}}^{y}\frac{1}{\sqrt{\frac{2}{m}\left(V(x)-V(x_{s})\right)}}dx.\label{eq:Period}
\end{equation}
The first integral can be re-expressed as
\begin{align}
{\cal T}= & 2\int_{x_{b}}^{x_{r}}\frac{1}{\sqrt{\frac{2}{m}\left(V(x)-V(x_{b})\right)}}-\frac{1}{\sqrt{\frac{2}{m}\left(V^{\prime}(x_{b})(x-x_{b})+\frac{1}{2}V^{\prime\prime}(x_{b})(x-x_{b})^{2}\right)}}dx\nonumber \\
 & +2\int_{x_{b}}^{x_{r}}\frac{1}{\sqrt{\frac{2}{m}\left(V^{\prime}(x_{b})(x-x_{b})+\frac{1}{2}V^{\prime\prime}(x_{b})(x-x_{b})^{2}\right)}}dx.
\end{align}
The additional integral can be conducted as
\begin{gather*}
\sim2\sqrt{\frac{m}{V^{\prime\prime}(x_{b})}}\ln\frac{\left|x_{r}-x_{b}+\frac{V^{\prime}(x_{b})}{V^{\prime\prime}(x_{b})}+\sqrt{\left(x_{r}-x_{b}+\frac{V^{\prime}(x_{b})}{V^{\prime\prime}(x_{b})}\right)^{2}-\left[\frac{V^{\prime}(x_{b})}{V^{\prime\prime}(x_{b})}\right]^{2}}\right|}{\left|\frac{V^{\prime}(x_{b})}{V^{\prime\prime}(x_{b})}\right|}.
\end{gather*}
Consequently, the following relation can be obtained
\begin{align}
 & \left|V^{\prime}(x_{b})\right|e^{\sqrt{\frac{V^{\prime\prime}(x_{b})}{m}}\frac{{\cal T}}{2}}\nonumber \\
= & \left|V^{\prime\prime}(x_{b})\right|\left|x_{r}-x_{b}+\frac{V^{\prime}(x_{b})}{V^{\prime\prime}(x_{b})}+\sqrt{\left(x_{r}-x_{b}+\frac{V^{\prime}(x_{b})}{V^{\prime\prime}(x_{b})}\right)^{2}-\left[\frac{V^{\prime}(x_{b})}{V^{\prime\prime}(x_{b})}\right]^{2}}\right|\nonumber \\
 & \qquad\qquad\cdot e^{\sqrt{\frac{V^{\prime\prime}(x_{b})}{m}}\int_{x_{b}}^{x_{r}}\frac{1}{\sqrt{\frac{2}{m}\left(V(\bar{x})-V(x_{b})\right)}}-\frac{1}{\sqrt{\frac{2}{m}\left(V^{\prime}(x_{b})(x-x_{b})+\frac{1}{2}V^{\prime\prime}(x_{b})(x-x_{b})^{2}\right)}}d\bar{x}}.\label{eq:V'(xb)exp(beta)}
\end{align}
Similarly, the relation between $x_{s}$ and $y$ can be obtained
as
\begin{align}
 & \left|V^{\prime}(x_{s})\right|e^{\sqrt{\frac{V^{\prime\prime}(x_{s})}{m}}\frac{{\cal T}}{2}}\nonumber \\
= & \left|V^{\prime\prime}(x_{s})\right|\left|y-x_{s}+\frac{V^{\prime}(x_{s})}{V^{\prime\prime}(x_{s})}+\sqrt{\left(y-x_{s}+\frac{V^{\prime}(x_{s})}{V^{\prime\prime}(x_{s})}\right)^{2}-\left[\frac{V^{\prime}(x_{s})}{V^{\prime\prime}(x_{s})}\right]^{2}}\right|\nonumber \\
 & \qquad\cdot e^{\sqrt{\frac{V^{\prime\prime}(x_{s})}{m}}\int_{x_{s}}^{y}\frac{1}{\sqrt{\frac{2}{m}\left(V(\bar{x})-V(x_{s})\right)}}-\frac{1}{\sqrt{\frac{2}{m}\left(V^{\prime}(x_{s})(x-x_{s})+\frac{1}{2}V^{\prime\prime}(x_{s})(x-x_{s})^{2}\right)}}d\bar{x}}.
\end{align}
Then, we can get
\begin{equation}
\lim_{x_{s}\leq x_{b}\rightarrow x_{F}}\frac{V^{\prime}(x_{b})}{V^{\prime}(x_{s})}=\frac{\left|x_{R}-x_{F}\right|e^{\omega_{F}\int_{x_{F}}^{x_{R}}\frac{1}{\sqrt{\frac{2}{m}\left(V(\bar{x})-V(x_{F})\right)}}-\frac{1}{\omega_{F}|x-x_{F}|}d\bar{x}}}{\left|y-x_{F}\right|e^{\omega_{F}\int_{x_{F}}^{y}\frac{1}{\sqrt{\frac{2}{m}\left(V(\bar{x})-V(x_{s})\right)}}-\frac{1}{\omega_{F}|x-x_{F}|}d\bar{x}}}=e^{\int_{y}^{x_{R}}\frac{\omega_{F}}{\sqrt{\frac{2}{m}\left(V(\bar{x})-V(x_{F})\right)}}d\bar{x}}.
\end{equation}

As for $V^{\prime}(x_{s})\frac{dy}{dx_{s}}$, $\frac{dy}{dx_{s}}$
can be expressed as
\begin{equation}
\frac{dy}{dx_{s}}=-\frac{\partial_{x_{s}}{\cal T}}{\partial_{y}{\cal T}}=\frac{1}{2}\sqrt{V(y)-V(x_{s})}\int_{x_{s}}^{y}\frac{V^{\prime}(x)-V^{\prime}(x_{s})}{\left[V(x)-V(x_{s})\right]^{\frac{3}{2}}}dx+1
\end{equation}
by using the general formula for derivative of the implicit function.
Therefore, 
\begin{align}
\lim_{{\cal T}\rightarrow\infty}V^{\prime}(x_{s})\frac{dy}{dx_{s}} & =\lim_{x_{s}\rightarrow x_{F}}V^{\prime}(x_{s})\left(\frac{1}{2}\sqrt{V(y)-V(x_{s})}\int_{x_{s}}^{y}\frac{V^{\prime}(x)-V^{\prime}(x_{s})}{\left[V(x)-V(x_{s})\right]^{\frac{3}{2}}}dx+1\right)\\
 & =\lim_{x_{s}\rightarrow x_{F}}V^{\prime}(x_{s})\sqrt{\frac{V(y)-V(x_{F})}{V(x_{s})-V(x_{F})}}\\
 & =\sqrt{2V^{\prime\prime}(x_{F})}\sqrt{V(y)-V(x_{F})},
\end{align}
where we have used 
\begin{equation}
\lim_{x_{s}\rightarrow x_{F}}\frac{\sqrt{V(x_{s})-V(x_{F})}}{V^{\prime}(x_{s})}=\sqrt{\frac{1}{2V^{\prime\prime}(x_{F})}}.
\end{equation}

The limit of $V^{\prime}(x_{b})\frac{d{\cal T}}{dx_{b}}$ can be calculated
from the limit of $\frac{{\cal T}}{\ln V^{\prime}(x_{b})}$ since
\begin{equation}
\lim_{x_{b}\rightarrow x_{F}}\frac{{\cal T}}{\ln V^{\prime}(x_{b})}=\lim_{x_{b}\rightarrow x_{F}}\frac{V^{\prime}(x_{b})}{V^{\prime\prime}(x_{b})}\frac{d{\cal T}}{dx_{b}}.
\end{equation}
And the limit of $\frac{{\cal T}}{\ln V^{\prime}(x_{b})}$ is already
known from Eq.(\ref{eq:V'(xb)exp(beta)}) as
\begin{equation}
\lim_{x_{b}\rightarrow x_{F}}\frac{{\cal T}}{\ln V^{\prime}(x_{b})}=-2\sqrt{\frac{m}{V^{\prime\prime}(x_{F})}}.
\end{equation}
As a result,
\begin{equation}
\lim_{x_{b}\rightarrow x_{F}}V^{\prime}(x_{b})\frac{d{\cal T}}{dx_{b}}=-2\sqrt{mV^{\prime\prime}(x_{F})}.
\end{equation}

By combining the result obtained above, the final result of the limit
of the ratio becomes
\begin{equation}
\lim_{{\cal T}\rightarrow\infty}\frac{\det\left[-\partial_{\tau}^{2}+\frac{1}{m}V^{\prime\prime}(x_{sF}^{{\cal T}}(\tau))\right]}{\det\left[-\partial_{\tau}^{2}+\frac{1}{m}V^{\prime\prime}(x_{sB}^{{\cal T}}(\tau;y))\right]}=-e^{2\int_{y}^{x_{R}}\frac{\omega_{F}}{\sqrt{\frac{2}{m}\left(V(x)-V(x_{F})\right)}}dx}.
\end{equation}
Therefore, 
\begin{equation}
\im E_{0}\approx\lim_{{\cal T}\rightarrow\infty}-\frac{\hbar}{2}\frac{1}{{\cal T}}e^{\int_{y}^{x_{R}}\frac{\omega_{F}}{\sqrt{\frac{2}{m}\left(V(x)-V(x_{F})\right)}}dx}e^{-\frac{1}{\hbar}S_{B}},
\end{equation}
where we have set $V(x_{F})=0$ to make $S_{sF}^{{\cal T}=\infty}=0$.
The decay rate at zero temperature becomes
\begin{equation}
\Gamma=\lim_{{\cal T}\rightarrow\infty}\frac{1}{{\cal T}}e^{\int_{y}^{x_{R}}\frac{\omega_{F}}{\sqrt{\frac{2}{m}\left(V(x)-V(x_{F})\right)}}dx}e^{-\frac{1}{\hbar}S_{B}}.
\end{equation}

When ${\cal T}\rightarrow\infty$, $y$ can be chosen to be $x_{F}$
in order to obtain a finite result. However, $y$ is selected artificially.
The final result depends on the approaching behavior of $y$ and is
not single. For comparison, we can re-express the exponent part as
\begin{equation}
e^{\int_{y}^{x_{R}}\frac{\omega_{F}}{\sqrt{\frac{2}{m}\left(V(x)-V(x_{F})\right)}}dx}=\frac{x_{R}-x_{F}}{y-x_{F}}e^{\int_{y}^{x_{R}}\frac{\omega_{F}}{\sqrt{\frac{2}{m}\left(V(x)-V(x_{F})\right)}}-\frac{1}{|x-x_{F}|}dx}\equiv\frac{1}{y-x_{F}}\frac{1}{\omega_{F}}A(y)\sqrt{S_{B}},
\end{equation}
where 
\begin{equation}
A(y)=\frac{\omega_{F}}{\sqrt{S_{B}}}(x_{R}-x_{F})e^{\int_{y}^{x_{R}}\frac{\omega_{F}}{\sqrt{\frac{2}{m}\left(V(\bar{x})-V(x_{F})\right)}}-\frac{1}{|x-x_{F}|}d\bar{x}}.
\end{equation}
Then, $\Gamma$ becomes
\begin{equation}
\Gamma=\lim_{{\cal T}\rightarrow\infty}\frac{1}{{\cal T}}\frac{1}{y-x_{F}}\frac{1}{\omega_{F}}A(y)\sqrt{S_{B}}e^{-\frac{1}{\hbar}S_{B}}=\lim_{{\cal T}\rightarrow\infty}\frac{1}{{\cal T}}\frac{1}{y-x_{F}}\frac{1}{\omega_{F}}\sqrt{\frac{\pi\hbar}{m\omega_{F}}}\frac{A(y)}{A(x_{F})}\Gamma_{\textrm{C\&C}}
\end{equation}
When $y$ is chosen to be $x_{F}$, $\frac{1}{y-x_{F}}$ becomes
infinite. Therefore, $\frac{1}{y-x_{F}}$ should reflect the zero
mode related to the bounce.

In order to obtain a finite final result, 
\begin{equation}
\lim_{{\cal T}\rightarrow\infty}y-x_{F}\sim\frac{1}{{\cal T}}.
\end{equation}
The choice is not unique. Furthermore, from Eq.(\ref{eq:V'(xb)exp(beta)})
we can know that
\begin{equation}
\lim_{{\cal T}\rightarrow\infty}V^{\prime}(x_{b})\sim x_{b}-x_{F}\sim e^{-\frac{\omega_{F}{\cal T}}{2}}.
\end{equation}
Therefore, it can be speculated that
\begin{equation}
\lim_{{\cal T}\rightarrow\infty}\frac{y-x_{F}}{x_{b}-x_{F}}\sim\infty,
\end{equation}
 if $y-x_{F}\sim\frac{1}{{\cal T}}$ was choose. Consequently, $y>x_{b}$,
which is expected. 

\begin{figure}[h]
\begin{centering}
\includegraphics[width=0.4\linewidth]{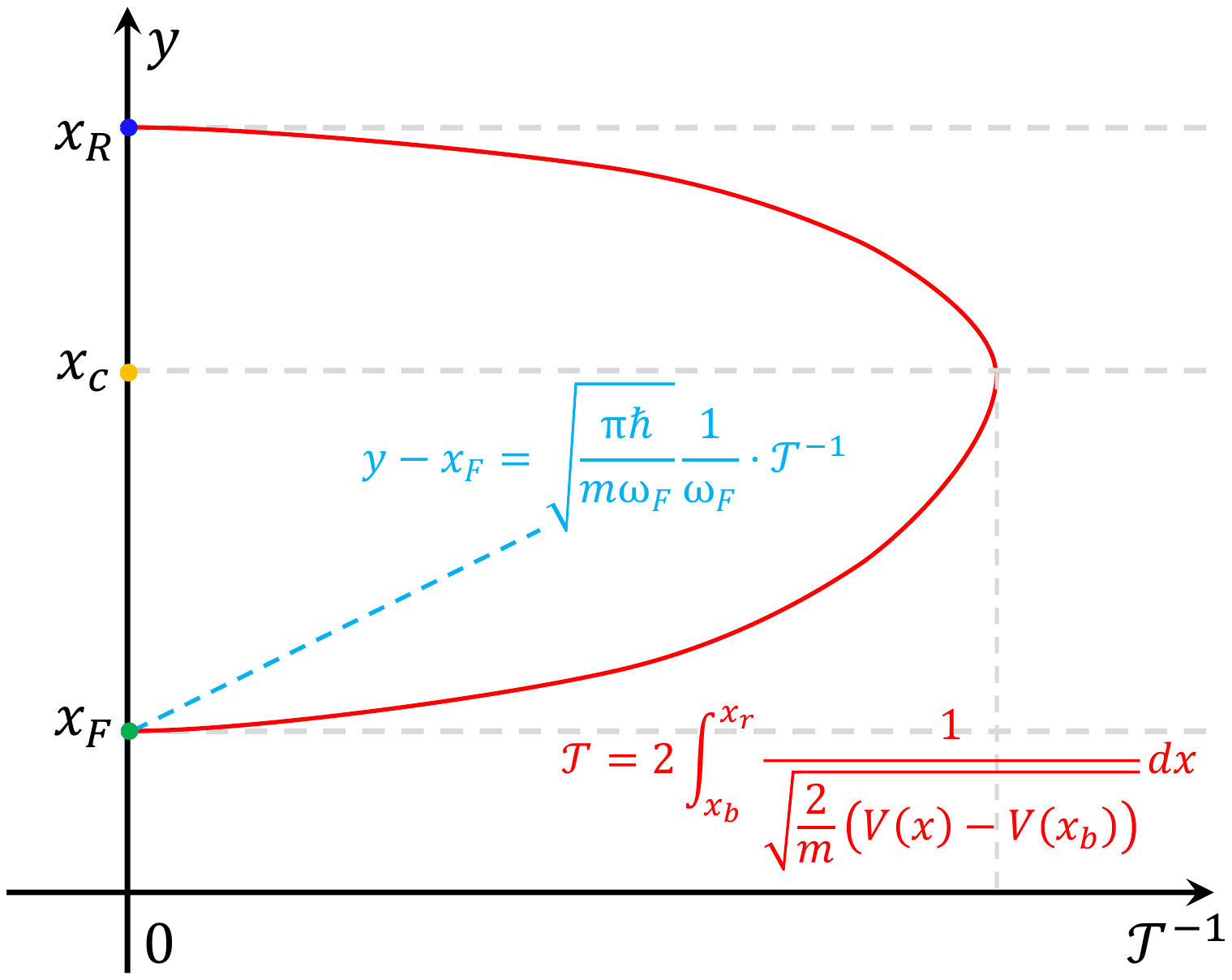}
\par\end{centering}
\caption{Zero Temperature Limit\label{fig:Zero-Temperature-Limit}}
\end{figure}

The final result of Callan and Coleman can be obtained by choosing
$y$ and ${\cal T}$ to be
\begin{equation}
\lim_{{\cal T}\rightarrow\infty}y-x_{F}=\sqrt{\frac{\pi\hbar}{m\omega_{F}}}\frac{1}{\omega_{F}{\cal T}}.
\end{equation}
However, the constant of proportionality between $y-x_{F}$ and $\frac{1}{{\cal T}}$
can be chosen arbitrarily (see Figure.\ref{fig:Zero-Temperature-Limit}).
Consequently, the result can be any value. This appears like the ratio
between two irrelevant infinite values. We believe that this fact
indicates the relation between the zero mode of the bounce solution
and the infinite time interval.

\section{Summary\label{sec:summary-and-discussion}}

In this paper we discussed two issues in the calculation of the decay
rate at zero temperature. We believe that the first problem related
to the appearance of the imaginary part can be solved once the states
in the rigged Hilbert space are taken into consideration. As for the
second problem, which is related to the timing to take the time interval
to infinity and is the main topic in our paper, we have recalculated
the transition amplitude by using saddle point approximation at finite
time interval. We have utilized the shot-F solution and the shifted-bounce
solution instead of the bounce solution. Since no zero eigenvalue
appears, there is no need to introduce the collective coordinate method
and the equivalence problem between the two infinite values is avoided.
When taking the time interval to infinity, we can obtain a finite
result if we choose the proper classical solutions. However, we find
that the final result depends on how the limit is taken and is not
unique. As we mentioned in Section.\ref{sec:Reconsideration}, the
validity of the collective coordinate method used to deal with the
zero mode as well as the equivalence between the integral over $c_{1}$
and $\tau_{c}$ should be discussed in more detail.

\section{Appendix}

\subsection{The Calculation of Functional Determinants}

The problem of deriving a functional determinant can be transformed
to solving the eigenequation under the corresponding initial conditions.
We here simply present the result (see \citep{Gelfand1960,McKane1995,Kirsten2003,Kirsten2004,Dunne2006,Dunne2008,Kirsten2010,Falco2017,Ossipov2018}
for more details).

The ratio of two functional determinant $\det\left[-\partial_{\tau}^{2}+W^{(i)}(\tau)\right]$
defined over $\left[-\frac{{\cal T}}{2},\frac{{\cal T}}{2}\right]$
by Dirichlet boundary conditions $\varphi_{n}^{(i)}(-\frac{{\cal T}}{2})=\varphi_{n}^{(i)}(\frac{{\cal T}}{2})=0$
can be expressed as
\begin{equation}
\frac{\det\left[-\partial_{\tau}^{2}+W^{(1)}(\tau)\right]}{\det\left[-\partial_{\tau}^{2}+W^{(2)}(\tau)\right]}=\frac{u^{(1)}(\frac{{\cal T}}{2})}{u^{(2)}(\frac{{\cal T}}{2})},
\end{equation}
where $u^{(i)}(\tau)$ is the solution to the following equation
\begin{equation}
\left[-\partial_{\tau}^{2}+W^{(i)}(\tau)\right]u^{(i)}(\tau)=0\label{eq:eigenfunction with lambda =00003D 0}
\end{equation}
and satisfies the following initial conditions
\[
\left(\begin{array}{c}
u^{(i)}(-\frac{{\cal T}}{2})\\
\dot{u}^{(i)}(-\frac{{\cal T}}{2})
\end{array}\right)=\left(\begin{array}{c}
0\\
1
\end{array}\right).
\]

Now, consider the fluctuation determinant $\det\left[-\partial_{\tau}^{2}+\frac{1}{m}V^{\prime\prime}(\bar{x})\right]$
of a classical solution $\bar{x}$. Since $\bar{x}$ is a classical
solution, it must obey the equation of motion
\begin{equation}
-\ddot{\bar{x}}+\frac{1}{m}V^{\prime}(\bar{x})=0.\label{eq:EoM}
\end{equation}
We can take the partial derivative of the equation of motion with
respect to a parameter $\alpha$ which is explicitly contained in
$\bar{x}(\tau)$ but not in $V(\bar{x})$ on both side. Then we get
\begin{equation}
\left[-\partial_{\tau}^{2}+\frac{1}{m}V^{\prime\prime}(\bar{x})\right]\partial_{\alpha}\bar{x}=0.
\end{equation}
So $\partial_{\alpha}\bar{x}$ is just a solution to Eq.(\ref{eq:eigenfunction with lambda =00003D 0}).
But pay attention, we can't assert that $\partial_{\alpha}\bar{x}$
is an eigenfunction since $\partial_{\alpha}\bar{x}$ may not satisfy
the boundary condition. 

When $\alpha$ is chosen to be $\tau$, $\partial_{\alpha}\bar{x}$
becomes $\dot{\bar{x}}(\tau)$, which is the velocity of the classical
solution $\bar{x}(\tau)$. If it is noticed that classical solutions
exist for every energy in a given range, then it is natural to realize
that $\bar{x}$ also depends on the energy $E$. Therefore, $\alpha$
can be chosen to be $E$ as well (see \citep{Marino2015}). Denote
$\nu_{\bar{x}}(\tau)\equiv m\partial_{E}\bar{x}$, then $\nu_{\bar{x}}(\tau)$
is another solution.

Before focusing on the function $u(\tau)$ that meets the initial
conditions, we would like to discuss the properties of $\nu_{\bar{x}}(\tau)$
in more detail.

First, the linear independence of $\dot{\bar{x}}(\tau)$ and $\nu_{\bar{x}}(\tau)$
can be proven immediately. Since $\bar{x}$ is the classical solution
in Euclidean time, its energy is conserved over Euclidean time $\tau$
: 
\begin{equation}
\frac{1}{2}m\dot{\bar{x}}^{2}-V(\bar{x})=E.
\end{equation}
Taking the derivative of Eq.(\ref{eq:EoM}) over energy $E$ gives
\begin{equation}
m\dot{\bar{x}}\partial_{E}\dot{\bar{x}}-V^{\prime}(\bar{x})\partial_{E}\bar{x}=1.
\end{equation}
Using the equation of motion and the commutative property of partial
derivatives, the following relation can be derived: 
\begin{equation}
\dot{\bar{x}}(\tau)\dot{\nu}_{\bar{x}}(\tau)-\ddot{\bar{x}}(\tau)\nu_{\bar{x}}(\tau)=\left|\begin{array}{cc}
\dot{\bar{x}}(\tau) & \nu_{\bar{x}}(\tau)\\
\ddot{\bar{x}}(\tau) & \dot{\nu}_{\bar{x}}(\tau)
\end{array}\right|=1.\label{eq:Wronskian of nu and xdot}
\end{equation}
Therefore, the Wronskian of $\dot{\bar{x}}(\tau)$ and $\nu_{\bar{x}}(\tau)$
is a non-zero constant, which indicates that $\dot{\bar{x}}$ and
$\nu_{\bar{x}}(\tau)$ are linearly independent.

Next, it must be emphasized that the periodicity of $\nu_{\bar{x}}(\tau)$
is different from that of $\bar{x}(\tau)$. Even if $\bar{x}(\tau)$
is the periodic function with period of $P$ ($P$ doesn't need to
be equal to ${\cal T}$), $\nu_{\bar{x}}(\tau)$ may not be periodic
since the period $P$ could also depend on the energy of classical
orbit. As a result, considering $\bar{x}(\tau)=\bar{x}(\tau+P)$,
the following relation holds
\begin{equation}
\nu_{\bar{x}}(\tau)=\nu_{\bar{x}}(\tau+P)+m\dot{\bar{x}}(\tau+P)\frac{dP}{dE}.\label{eq:v(t)-v(t+P)}
\end{equation}

Finally, the parity of $\nu_{\bar{x}}(\tau)$ is the same as that
of $\bar{x}(\tau)$. If $\bar{x}(\tau)$ is an odd or even function
with regard to the Euclidean time $\tau$, then $\nu_{\bar{x}}(\tau)$
will exhibit the same parity.

Since two independent solutions have been found, $u(\tau)$ can be
constructed as
\begin{equation}
u(\tau)=\dot{\bar{x}}(-\frac{{\cal T}}{2})\nu_{\bar{x}}(\tau)-\dot{\bar{x}}(\tau)\nu_{\bar{x}}(-\frac{{\cal T}}{2}).\label{eq:u2}
\end{equation}

\begin{enumerate}
\item For a shot solution : 
\begin{equation}
u_{S}(\tau)=\dot{x}_{S}^{{\cal T}}(-\frac{{\cal T}}{2})\nu_{S}^{{\cal T}}(\tau)-\dot{x}_{S}^{{\cal T}}(\tau)\nu_{S}^{{\cal T}}(-\frac{{\cal T}}{2}).
\end{equation}
$u_{S}(\frac{{\cal T}}{2})$ becomes 
\begin{align}
u_{S}(\frac{{\cal T}}{2}) & =\dot{x}_{S}^{{\cal T}}(-\frac{{\cal T}}{2})\nu_{S}^{{\cal T}}(\frac{{\cal T}}{2})-\dot{x}_{S}^{{\cal T}}(\frac{{\cal T}}{2})\nu_{S}^{{\cal T}}(-\frac{{\cal T}}{2})\\
 & =-\left[\dot{x}_{S}^{{\cal T}}(\tau)\nu_{S}^{{\cal T}}(\tau)\right]_{-\frac{{\cal T}}{2}}^{\frac{{\cal T}}{2}}\\
 & =-\frac{\partial S_{S}}{\partial E_{S}}.
\end{align}
\item For the shifted-bounce solution :
\begin{equation}
u_{sB}(\tau)=\dot{x}_{sB}^{{\cal T}}(-\frac{{\cal T}}{2})\nu_{sB}^{{\cal T}}(\tau)-\dot{x}_{sB}^{{\cal T}}(\tau)\nu_{sB}^{{\cal T}}(-\frac{{\cal T}}{2})
\end{equation}
$u_{sB}(\frac{{\cal T}}{2})$ becomes 
\begin{align}
u_{sB}(\frac{{\cal T}}{2}) & =\dot{x}_{sB}^{{\cal T}}(-\frac{{\cal T}}{2})\nu_{sB}^{{\cal T}}(\frac{{\cal T}}{2})-\dot{x}_{sB}^{{\cal T}}(\frac{{\cal T}}{2})\nu_{sB}^{{\cal T}}(-\frac{{\cal T}}{2})\\
 & =\dot{x}_{sB}^{{\cal T}}(\frac{{\cal T}}{2})\left[\nu_{sB}^{{\cal T}}(\frac{{\cal T}}{2})-\nu_{sB}^{{\cal T}}(-\frac{{\cal T}}{2})\right]\\
 & =-m\left[\dot{x}_{sB}^{{\cal T}}(\frac{{\cal T}}{2})\right]^{2}\frac{d{\cal T}}{dE_{B}}.
\end{align}
\end{enumerate}
Therefore, the following relation can be obtained.
\begin{equation}
\frac{\det\left[-\partial_{\tau}^{2}+\frac{1}{m}V^{\prime\prime}(x_{sF}^{{\cal T}}(\tau))\right]}{\det\left[-\partial_{\tau}^{2}+\frac{1}{m}V^{\prime\prime}(x_{sB}^{{\cal T}}(\tau))\right]}=-\frac{-\frac{\partial S_{sF}}{\partial E_{sF}}}{m\left[\dot{x}_{sB}^{{\cal T}}(\frac{{\cal T}}{2})\right]^{2}\frac{d{\cal T}}{dE_{B}}}.
\end{equation}

\begin{acknowledgments}
The authors is grateful to Prof. Koji Harada and Prof. Shuichiro Tao for valuable discussions. This work is supported by Kyushu University Leading Human Resources Development Fellowship Program (Quantum Science Area).
\end{acknowledgments}

\newpage{}

\bibliographystyle{plunsrt}
\bibliography{reference}

\end{document}